\documentclass[a4paper,fleqn,usenatbib]{mnras}

\usepackage{newtxtext,newtxmath}

\usepackage[T1]{fontenc}
\usepackage{ae,aecompl}

\usepackage{graphicx}	
\usepackage{amsmath}	
\usepackage{amssymb}	
\usepackage{xcolor}
\usepackage{multirow}

\newcommand{\kms}{\,km\,s$^{-1}$}
\newcommand{\sig}{$\sigma$\space}
\newcommand{\mgf}{$\rm Mg4780$\space}
\newcommand{\tioi}{$\rm TiO_{1}$\space}
\newcommand{\tioiio}{$\rm TiO_{2_{SDSS}}$\space}
\newcommand{\tioii}{$\rm TiO_{2}$}
\newcommand{\naii}{$\rm Na8190_{SDSS}$\space}

\newcommand{\caii}{$\rm Ca2$}

\newcommand{\mgb}{$\rm Mgb$}
\newcommand{\afe}{$[\alpha/{\rm Fe}]$}
\newcommand{\mgfep}{$\rm [MgFe]'$}

\newcommand{\hdf}{$\rm H_{\delta_{F}}$}

\title[Comparing IMF-sensitive indices of intermediate-mass quiescent galaxies in various environments]{Comparing IMF-sensitive indices of intermediate-mass quiescent galaxies in various environments}

\author[E. Eftekhari et al.]{
Elham Eftekhari,$^{1,2}$\thanks{E-mail: elhamea@iac.es (EE)}
Moein Mosleh,$^{3,4}$
Alexandre Vazdekis$^{1,2}$
and Saeed Tavasoli$^{5}$
\\
$^{1}$Instituto de Astrof\'isica de Canarias, E-38200 La Laguna, Tenerife, Spain\\
$^{2}$Departamento de Astrof\'isica, Universidad de La Laguna, E-38205 La Laguna, Tenerife, Spain\\
$^{3}$Biruni Observatory, Shiraz University, Shiraz 71454, Iran\\
$^{4}$Physics Department, Shiraz University, Shiraz 71454, Iran\\
$^{5}$Faculty of Physics, Kharazmi University, Mofateh Ave., Tehran, Iran
}

\date{Accepted XXX. Received YYY; in original form ZZZ}

\pubyear{2018}

\begin{document}
\label{firstpage}
\pagerange{\pageref{firstpage}--\pageref{lastpage}}
\maketitle
\begin{abstract}
Using samples drawn from the Sloan Digital Sky Survey, we study for the first time the relation between large-scale environments (Clusters, Groups and Voids) and the stellar Initial Mass Function (IMF). We perform an observational approach based on the comparison of IMF-sensitive indices of quiescent galaxies with similar mass in varying environments. These galaxies are selected within a narrow redshift interval ($ 0.020 < z < 0.055 $) and spanning a range in velocity dispersion from 100 to 200\kms. The results of this paper are based upon analysis of composite spectra created by stacking the spectra of galaxies, binned by their velocity dispersion and redshift. The trends of spectral indices as measured from the stacked spectra, with respect to velocity dispersion, are compared in different environments. We find a lack of dependence of the IMF on the environment for intermediate-mass galaxy regime. We verify this finding by providing a more quantitative measurement of the IMF variations among galactic environments using MILES stellar population models with a precision of $\Delta\Gamma_{b}\sim0.2$.

\end{abstract}




\section{INTRODUCTION}

 \vspace{3 mm}

A widely adopted approach to constrain the formation and evolution of quenched galaxies is to analyze their stellar population  properties (e.g. stellar mass, age and star-formation histories) over cosmic time and as a function of different environments. The stellar Initial Mass Function (IMF), which gives the mass spectrum of stars at birth, is one of the key ingredients in this puzzle.

Traditionally the IMF has often been considered as universal across galaxy types and cosmic time, mostly because of a lack of evidence of IMF variations among stellar clusters and OB associations in the Milky Way \citep{kroupa2002, chabrier2003, bastian2010}. However, there is a growing body of publications showing a systematic variation of the IMF with the velocity dispersion ($\sigma$) of early-type galaxies (ETGs) (e.g. \citet{cappellari2012, la2013, ferreras2013, spiniello2014}). Moreover, recent works by several groups have reported radial variation of the IMF within massive ETGs \citep{martin2015a, martin2015b, la2016, davis2016, van2017, sarzi2018}. 

There are three independent probes of the IMF in the unresolved stellar populations: 1) strong gravitational lenses, 2) stellar kinematics and 3) IMF-sensitive spectroscopic features. By employing the first two probes, the total mass distribution of galaxies can be obtained (dark matter + stellar mass +  mass of gas). With a few assumptions about dark matter profile and also considering that the stellar mass density follows the luminosity density, it is possible to separate stellar and dark matter components of galaxies. This leads to a constraint on the stellar mass-to-light ratio (M/L) and consequently a constraint on the IMF \citep{ferreras2008, ferreras2010, auger2010, treu2010, thomas2011, cappellari2012, cappellari2013, conroy2013b, tortora2013a, smith2013, tortora2014, posacki2014, tortora2016, leier2016}.

The spectroscopic technique is the most direct method to constrain the IMF, with respect to other techniques and is most sensitive to the ratio between dwarf and giant stars. Several IMF-sensitive spectral features are known as a dwarf/giant discriminators. For example, \ion{Na}{i} 8190, Mg 4780 and TiO features reveal the presence of dwarfs in integrated light spectra while the calcium triplet, CaT, is strongest in the giants \citep{cohen1978, carter1986, schiavon1997, schiavon2000, cenarro2003, vazdekis2003, van2010, van2011, conroy2012, van2012, la2013, spiniello2014}. The study of these spectral features in massive local ETGs have shown that the strength of IMF-sensitive indices varies with ETG velocity dispersion. The IMF becomes progressively bottom-heavy, i.e. with a higher ratio of low- to high- mass stars than the Milky Way IMF, as velocity dispersion increases \citep{cenarro2003, van2010, ferreras2013, van2012, la2013, spiniello2014, lagattuta2017}. 

Despite the numerous studies on the IMF, very little is known about the dependence of stellar IMF on large-scale environment. Galaxies that reside in dense environments are more likely to experience interactions such as ram pressure stripping \citep{abadi1999, cen2014, bahe2015}, galaxy harassment \citep{bialas2015} and strangulation of gas from neighbors \citep{kawata2008} while galaxies in low-density environments remain largely untouched. So it is expected that these diverse evolutionary processes affect differently on the properties of galaxies. Correlations have been found between environmental density and galaxy morphology \citep{dressler1980, postman1984, postman2005, bamford2009}, color \citep{pimbblet2002, blanton2005}, Sersic index \citep{blanton2005}, concentration \citep{hashimoto1999}, star formation rate \citep{lewis2002, gomez2003, kauffmann2004, darvish2016}, age, metallicity \citep{kuntschner2002, thomas2005, sanchez2006}  and quenched galaxy fraction \citep{baldry2006, peng2012, darvish2016}. 

Here we study the relation of the IMF of galaxies in varying environment. In \citet{rosani2018}, the mass of the dark matter host halo of galaxies are used as a proxy for the environment. The authors did not find any dependency of the IMF slope on host halo mass. This motivates us to study the IMF in two distinct samples at extremely low-density environments of voids and walls of the cosmic web (groups and clusters). Most of the previous studies have focused on the IMF of galaxies in dense regions \citep{van2010, van2011, conroy2012b, smith2012, martin2015b, tortora2016, davis2016, zieleniewski2017, leier2016}. However, cosmic voids are excellent probes of the effect of environment on galaxy formation and evolution; complex processes such as close encounters and galaxy mergers are rare in void regions \citep{croton2008, vandeWeygaert2011, kreckel2014, tavasoli2015, moorman2016}. With the purpose of verifying the impact of the galactic environment on the stellar IMF, here  we compare the trend of IMF-sensitive spectral indices with \sig in the cluster, group and void regions. In this way, we avoid the well-known bias that more massive galaxies inhabit denser environments \citep{dressler1980}. The spectral indices, studied in the present work, are measured in stacked spectra obtained from Sloan Digital Sky Survey  (SDSS), Data Release 7 (DR7) \citep{abazajian2009}. In addition to our empirical approach, we perform a more quantitative assessment to support our conclusions.

There are two factors that make our study unique in the field of stellar IMF: first, we used isolated void galaxies which are less affected by mergers, second we used quiescent galaxies rather than red-sequence ones due to avoid the contribution of obscured star-forming galaxies. It is well known that dust-obscured star-forming galaxies can display colors coincident with truly passive populations on the red sequence. Therefore, selecting red galaxies based on a single color in a color-magnitude/mass diagram results in a selection of both passive galaxies and reddened star-forming galaxies. In recent years, it has been shown that galaxies with very little or no star-forming activity tend to be found in a particular region in color-color space \citep{patel2011}. Using the color-color diagram in this work, we separated dusty star-forming galaxies and truly quiescent ones.

This paper is structured as follows: In Section \ref{sec:data} we describe the data and sample selection; The procedure of stacking spectra and also the absorption line indices used in this work are presented in Section \ref{sec:method}. The main results are introduced in Section \ref{sec:results}, i.e. comparison of the trend of stellar population indicators in low and high dense environments (Section~\ref{sec:sigmatrend}), as well as a quantitative measure for the IMF variations (Section~\ref{sec:models}). A summary and a discussion are given in Section \ref{sec:discussion}. 

\section{DATA AND SAMPLE SELECTION} \label{sec:data}

The purpose of this study is to investigate the role of the background density on the IMF-sensitive indices of galaxies, especially at the extremely low-density environments of the voids. Hence, we have used the same sample of \citet{mosleh2018}. This sample consists of three sub-samples: isolated quiescent void galaxies, quiescent group galaxies and quiescent cluster galaxies. Isolated quenched void galaxies are extracted from a catalog of void galaxies by \citet{tavasoli2015} which is generated from a volume-limited spectroscopic sample of the Sloan Digital Sky Survey, Data Release 10 (SDSS DR10) \citep{ahn2014}. This catalog consists of 1014 void galaxies, brighter than $M_{r}=-19$ with the redshift range $0.01 < z < 0.055$. \citet{mosleh2018} divided galaxies in this catalog into star-forming and quiescent ones by using the \textit{urz} color criteria and quiescent galaxies are cross-matched with the groups catalog of \citet{tempel2014}. Then quenched void galaxies with more than one companion are removed which reduced the number of galaxies in void sub-sample to 147. \citet{mosleh2018} did so to ensure that galaxies in their sub-sample are isolated void galaxies, which magnify any variation between galaxy parameters at low- and high-density environments.

\begin{figure}
	\includegraphics[width=\linewidth, clip]{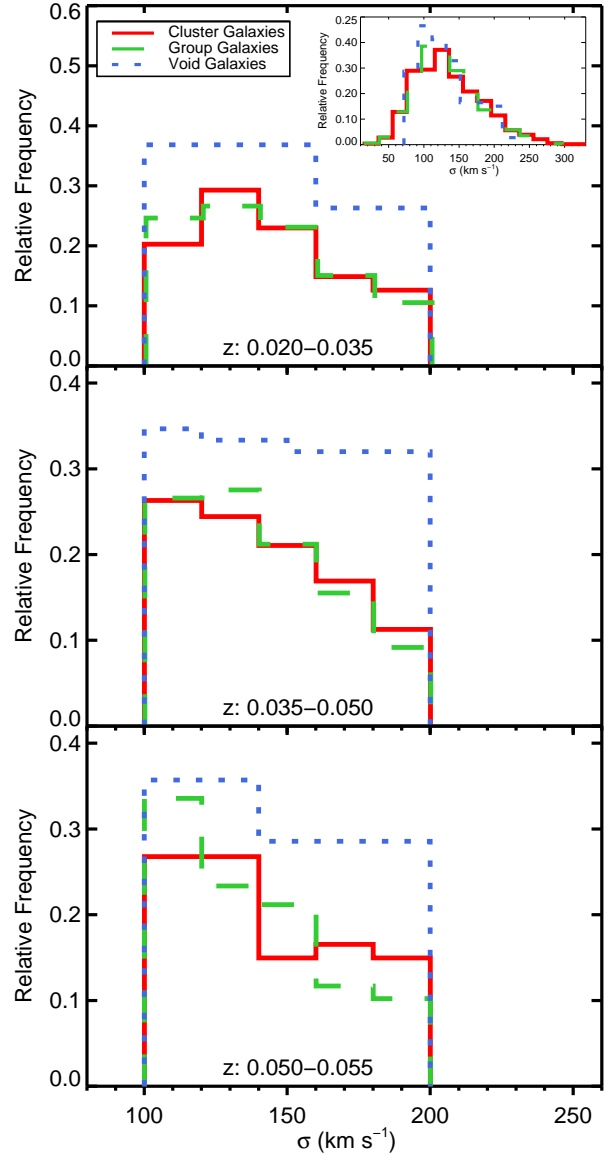}
	\caption{Histograms of galaxies in different redshift bins. The red histograms correspond to galaxies in clusters, the green dashed lines correspond to the histograms of group galaxies and the blue dotted lines represent histograms of void galaxies. The inset shows the distribution of the \sig of quiescent galaxies in clusters, groups and voids in \citet{mosleh2018} sample. It shows that although the  group and cluster samples are mass-matched with the void sample, their \sig distribution are also matched. }
	\label{fig:fig1}
\end{figure}

To compare the galaxies in low-density environments with those in intermediate and high-density regions, \citet{mosleh2018} selected two samples of galaxies in rich clusters (with more than 15 galaxies) and rich groups (with 4-8 spectroscopic members) randomly from the \citet{tempel2014} catalog. Again the \textit{urz} color criteria is employed to separate star-forming and quiescent galaxies. Quiescent cluster and group galaxies are selected in such a way that their stellar mass distributions are matched with the mass distribution of the isolated quiescent void sample but contain 7$\times$ more galaxies (see \citet{mosleh2018} for more details). Although the sub-samples of different environments are mass-matched, their velocity dispersion distributions are also similar. This is shown as an inset to the top panel of Figure~\ref{fig:fig1}.
 
We restricted our samples to the galaxies with redshifts higher than 0.02, because there are no isolated quiescent void galaxies with $z<0.02$ in this sample. Also, we only selected galaxies with \sig higher than 100 \kms \space and lower than 200 \kms (because below 100 \kms \space the sensitivity of IMF indices to IMF variations is too low and higher than 200 \kms \space the statistics of void galaxies is too poor). As the aperture of an SDSS spectroscopic fiber ($3\ensuremath {{}^{\prime \prime }}$) samples only the inner parts of nearby galaxies and this fraction is different at each redshift, to construct comparable samples, we split the galaxies in clusters, groups, and voids to three redshift bins: [0.020-0.035], [0.035-0.050], [0.050-0.055]. Then cluster and group galaxies are binned into five velocity dispersion intervals with a width of 20 \kms \space at each redshift bin. The result is 15 bins of galaxies in each cluster and group environments. Due to the low number of void galaxies in our sample, we requested three $\sigma$-bins at each redshift interval of void sub-sample. In total, there are 9 bins of void galaxies. In Figure~\ref{fig:fig1}, the histogram of galaxies in each environment has been shown for three redshift bins. Since the number of our cluster, group and void galaxies are maximum in redshifts 0.035 to 0.050, we mainly focus on this redshift-bin throughout the rest of this paper. The analysis of other redshift intervals can be found in the Appendix~\ref{app:app2}.

\section{METHODOLOGY} \label{sec:method}

Our study is based on comparing stellar IMF-sensitive indices of different environments. The SDSS DR7 spectral range allows us to study various of these features. As the SDSS data have low S/N, we stacked the spectra of galaxies to achieve a S/N that allows us to apply a detailed absorption line index comparisons. The cost is that we loss individual galaxy identities. To create the stacked spectra, for each bin we first measured the spectral indices of individual galaxies using the Indexf code\footnote{\url{http://indexf.readthedocs.io}}\citep{cardiel2010}. Before measuring the spectral indices, we prepared the spectra of galaxies: we masked bad pixels and smoothed each spectrum to $\sigma$=214\kms \space which is the spectral resolution, best suited for low- and intermediate-mass galaxies in the Line Index System (LIS) \citep{vazdekis2010}. We took into account the wavelength dependent SDSS resolution when smoothing each spectrum to $\sigma$=214\kms.  Also, as the SDSS data are stored in vacuum wavelength, we converted them to air wavelengths. Then the spectra are de-redshifted to the rest-frame. After running the Indexf code and measuring line-strengths of individual galaxies, we determined the median values of strengths of IMF-sensitive features in each bin and trimmed away outliers using $3\sigma$ cut-off. The excluded galaxies in each index measurement were flagged. If a galaxy was flagged for more than 1 spectral line, its spectrum was rejected in the analysis. About 2\% of galaxies in the cluster and group samples have been excluded from the original samples, but no galaxies in the voids have been excluded.  

\begin{figure*}
	
	\includegraphics[width=\linewidth, clip]{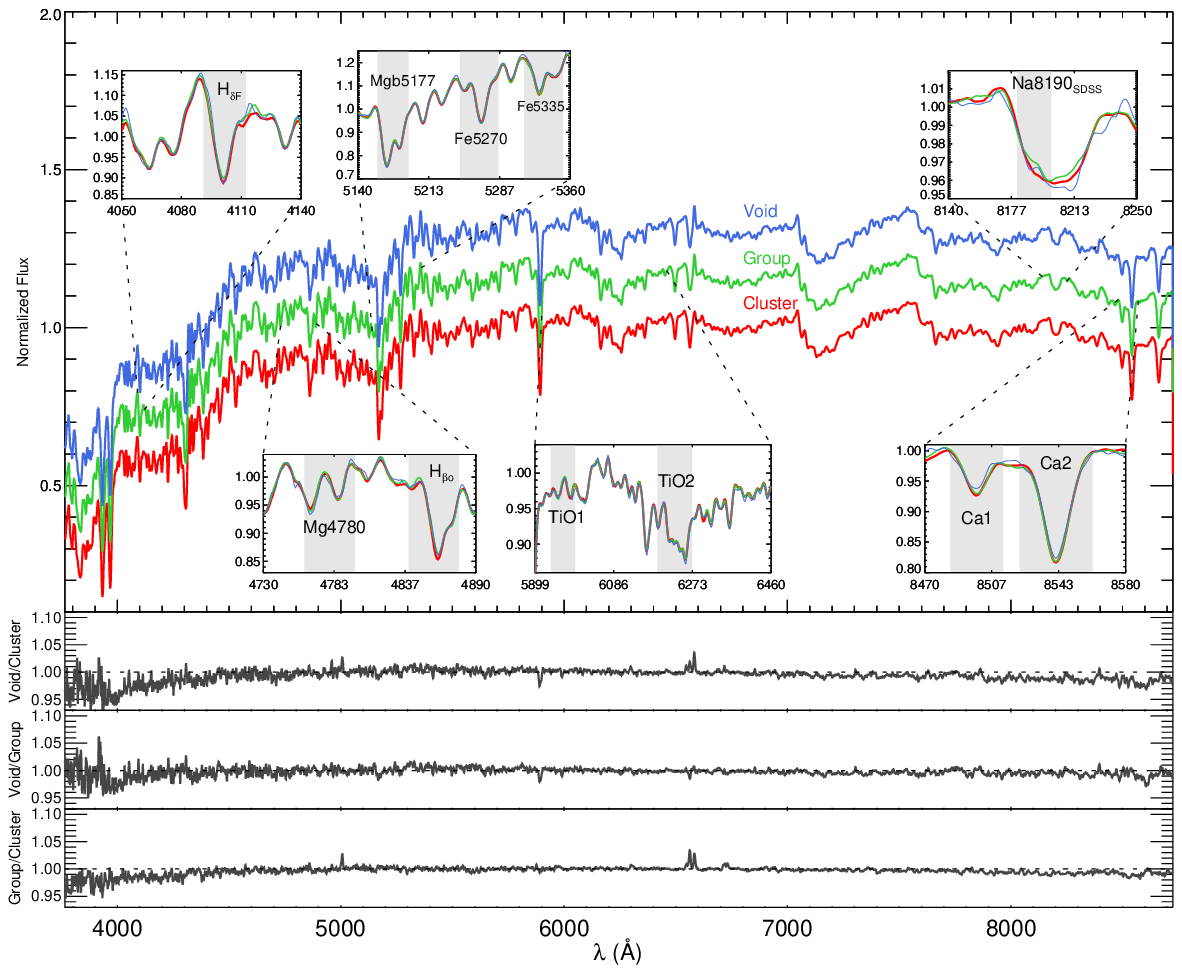}
	\caption{Upper panel: stacked spectra of cluster, group and void galaxies in the redshift-bin 0.035-0.050 and velocity dispersion-bin [150-200] \kms. For illustrative purpose, we stacked the spectra of group/cluster galaxies with velocity dispersions between 150 to 200 \kms \space to have a stacked spectrum comparable with void galaxies. Different colors correspond to different environments as labeled. The insets are zoomed-in versions of the spectral indices used in the present work. The stacked spectrum of group and void galaxies are shifted for display purpose. Lower panels: the ratio between the stacked spectra of voids and clusters, voids and groups and finally, groups and clusters are shown, respectively.}
	
	\label{fig:fig2}
	
\end{figure*}

After determining the final  sample, we created the stacked spectra in the following way: first, problematic pixels were masked. Then each individual spectrum has been convolved to $\sigma$=214\kms \space and corrected to air wavelengths. Next, we shifted the spectra into the rest-frame and binned them onto a common wavelength grid with a fixed dispersion of 1\AA \space in the range available for all galaxies, from 3760\AA \space to 8730\AA. Each individual spectrum was normalized to the median flux between 5000\AA \space and 8000\AA \space where the spectrum is relatively flat. This median is stored for use in the next step. When stacking the spectra, we multiplied each spectrum in each bin by the median of all individual medians in that bin. This ensures that the spectra of galaxies in each bin are in the same flux level. Finally, the stacked spectra are made by taking the median of fluxes at each wavelength. The error of the stacked spectra was estimated by bootstrap resampling which indicates the spread of
measurements across the sample of galaxies from which the stacks were created. Figure~\ref{fig:fig2} shows the stacked spectra of our samples in the redshift bin 0.035-0.050 for $\sigma$-bin of [150-200] \kms. Note that we made this bin for group/cluster galaxies to have the same $\sigma$-bin as in void galaxies, for illustrative purpose. We zoom into those regions used in the present work. In the lower panels of Figure~\ref{fig:fig2}, the stacked spectra of galaxies in different environments are divided by each other.

IMF-sensitive line strength indices were measured from the stacked spectra using the Indexf code. In Tables~\ref{tab:tab1}, \ref{tab:tab2} and \ref{tab:tab3} we report the number of stacked galaxy spectra with their S/N ratio, computed within the central bandpass of spectral indices in the Indexf code.

\begin{table*}
	\caption{Properties of the stacked spectra of cluster galaxies in bins of redshift and velocity dispersion}
	\begin{tabular}{c c c c c c c c}
		\hline
		$Redshift$ & $\sigma$ range & $N_{gal}$ & 
		\multicolumn{5}{c}{S/N [\footnotesize{$\mathring{A}^{-1}$]}}\\ 
		& $\rm [km/s]$ & \# & Ca2 & \naii & \tioi & \tioiio & \mgf \\
		(1) & (2) & (3) & (4) &(5) &(6) & (7) & (8) \\
		\hline
		\multirow{5}{*}{$0.020$--$0.035$}&$100$--$120$	&  $45$ &         121 &      147 &      416 &      545 &      104\\
		&	$120$--$140$& $65$ &      228 &      325 &      522 &      630 &      180\\
		&	$140$--$160$&  $51$ & 177 &      221 &      453 &      529 &      117\\
		&	$160$--$180$&  $33$ &       125 &      185 &      310 &      461 &      122\\
		&	$180$--$200$& $28$ &  140 &      175 &      333 &      379 &      129\\
		
		\hline
		\multirow{5}{*}{$0.035$--$0.050$}&	$100$--$120$&   $70$ &177 &      257 &      463 &      538 &      175\\
		&	$120$--$140$&  $65$ &	163 &      310 &      457 &      517 &      176\\
		&	$140$--$160$&  $56$ &	188 &      248 &      440 &      524 &      133\\
		
		&	$160$--$180$&   $45$ &	140 &      217 &      347 &      412 &      131\\
		&	$180$--$200$&  $30$ &	150 &      206 &      346 &      383 &       81\\
		
		\hline
		\multirow{5}{*}{$0.050$--$0.055$}&	$100$--$120$&  $34$ &	117 &      149 &      269 &      379 &       90\\
		&	$120$--$140$&  $34$ &106 &      119 &      305 &      366 &      108\\
		&	$140$--$160$&   $19$ &	67 &       88 &      172 &      259 &       58\\
		&	$160$--$180$&  $21$ &	101 &      165 &      259 &      297 &      122\\
		&	$180$--$200$& $19$ &	59 &       99 &      210 &      248 &       86\\
	
		\hline
	\end{tabular}
	\label{tab:tab1}
\end{table*}

\begin{table*}
	\caption{Properties of the stacked spectra of group galaxies in bins of redshift and velocity dispersion}
	\begin{tabular}{c c c c c c c c}
		\hline
		$Redshift$ & $\sigma$ range & $N_{gal}$ & 
		\multicolumn{5}{c}{S/N [\footnotesize{$\mathring{A}^{-1}$]}}\\ 
		& $\rm [km/s]$ & \# & Ca2 & \naii & \tioi & \tioiio & \mgf \\
		(1) & (2) & (3) & (4) &(5) &(6) & (7) & (8) \\
		\hline
		\multirow{5}{*}{$0.020$--$0.035$}&$100$--$120$	& $48$ &   191 &      259 &      440 &      461 &      128\\
		&	$120$--$140$& $53$ &    158 &      259 &      424 &      529 &      186\\
		&	$140$--$160$& $44$ &     157 &      305 &      382 &      449 &      106\\
		&	$160$--$180$&  $32$ &    143 &      176 &      295 &      393 &       73\\
		&	$180$--$200$&   $22$ & 132 &      135 &      263 &      338 &       81\\

		\hline
		\multirow{5}{*}{$0.035$--$0.050$}&	$100$--$120$&   $84$ &	186 &      288 &      558 &      630 &      183\\
		&	$120$--$140$&   $86$ &	192 &      322 &      498 &      620 &      145\\
		&	$140$--$160$& $67$ &	177 &      235 &      424 &      566 &      133\\
		
		&	$160$--$180$&  $50$ &	175 &      236 &      420 &      511 &      143\\
		&	$180$--$200$&  $29$ &	121 &      143 &      267 &      345 &       95\\

		\hline
		\multirow{5}{*}{$0.050$--$0.055$}&	$100$--$120$&  $46$ &	139 &      170 &      351 &      424 &      145\\
		&	$120$--$140$&  $32$ &	81 &      146 &      234 &      354 &       90\\
		&	$140$--$160$&    $29$ &	155 &      158 &      320 &      367 &      121\\
		&	$160$--$180$&   $16$ &	103 &      109 &      245 &      297 &       95\\
		&	$180$--$200$&  $14$ &	91 &      135 &      198 &      253 &       98\\
	
		\hline
	\end{tabular}
	\label{tab:tab2}
\end{table*}

\begin{table*}
	\caption{Properties of the stacked spectra of void galaxies in bins of redshift and velocity dispersion}
	\begin{tabular}{c c c c c c c c}
		\hline
		$Redshift$ & $\sigma$ range & $N_{gal}$ & 
		\multicolumn{5}{c}{S/N [\footnotesize{$\mathring{A}^{-1}$]}}\\ 
		& $\rm [km/s]$ & \# & Ca2 & \naii & \tioi & \tioiio & \mgf \\
		(1) & (2) & (3) & (4) &(5) &(6) & (7) & (8)    \\
		\hline
		
		\multirow{3}{*}{$0.020$--$0.035$}& $100$--$130$	&  $7$ &   93 &      124 &      161 &      184 &       55\\
		&	$130$--$160$&  $7$ &	      94 &      104 &      148 &      178 &       49
		\\
		&	$160$--$200$&  $5$ &	41 &       50 &      114 &      148 &       44\\
		
		\hline 
		
		\multirow{3}{*}{$0.035$--$0.050$}& $100$--$120$	&  $26$ &         96 &      129 &      254 &      340 &       76
		\\
		&	$120$--$150$&  $25$ &	     125 &      150 &      275 &      357 &      152
		\\
		&	$150$--$200$&  $24$ &	      78 &       87 &      182 &      259 &       51
		\\
		
		\hline 
		
		\multirow{3}{*}{$0.050$--$0.055$}&	$100$--$115$ &  $5$ &         56 &       62 &      131 &      175 &       40
		\\
		&	$115$--$140$&  $5$ &	      67 &      104 &      133 &      213 &       52
		\\
		&	$140$--$200$&  $4$ &	      72 &      116 &      162 &      236 &       86
		\\
		
		\hline 
		
	\end{tabular}
	\label{tab:tab3}
\end{table*}

We investigate the dependency of the age-, metallicity- and elemental abundance-sensitive indices on the environment in addition to IMF-sensitive indices. This is necessary to confirm that our findings of the IMF-sensitive spectral features are not caused by the environmental dependency of the other stellar population parameters. Then we perform a quantitative measurements of IMF variation among low and high density environments.

We summarize here the various absorption line indices used in this paper.

\subsection{IMF-sensitive indices}

Using near-infrared spectra of M-dwarf and M-giant stars, \citet{cohen1978} found that the near-IR \ion{Ca}{ii} triplet lines are strongest in the spectrum of giant stars while the \ion{Na}{i} feature is stronger in dwarfs and can be used as dwarf/giant indicators. Also, the studies of \citet{schiavon1997} and \citet{schiavon2000} on the spectra of cool stars and the integrated spectra of single-stellar populations \citep{vazdekis2012} confirmed that the Na I doublet  is an IMF-sensitive feature. Moreover, one of the most prominent molecular bands in M-dwarfs is TiO \citep{mould1975}. So the \tioi and \tioii \space indices which measure the absorption of TiO molecular band provide sensitive tracers of low-mass cool stars. In addition, exploring MILES extended Simple Stellar Populations (SSPs) \citep{vazdekis2012, ricciardelli2012} by \citet{la2013} and \citet{conroy2012} SSP models by  \citet{spiniello2012}, it was found that the index strength of \tioi and \tioii \space indices clearly increase with steeper IMF slopes. Also in a study by \citet{la2013}, it was found that \mgf has a significant dependence on the IMF slope. Hence, this index can be used as an IMF-sensitive feature. Table~\ref{tab:tab4} summarizes the definition of line indices, used in this paper for studying environmental dependency of the IMF. Note that instead of \ion{Ca}{ii} triplet, we used the Ca2 line. The reason for this choice is that Ca1 is too faint and Ca3 cannot be measured over the SDSS spectral range for most galaxies in our sample.

\begin{table*}
	\caption{Definition of IMF-sensitive indices used in this study. }
	\begin{tabular}{c c c c c c}
		\hline
		Index & Units & Blue Pseudo-continuum & Central feature & Red Pseudo-continuum & Source \\
		&       & [\AA] & [\AA] & [\AA] &  \\
		\hline
		\mgf    & \AA & $4738.900$--$4757.300$ & $4760.800$--$4798.800$ & $4819.800$--$4835.500$ & \citet{serven2005} \\
		\tioi   & mag & $5816.625$--$5849.125$ & $5936.625$--$5994.125$ & $6038.625$--$6103.625$& \citet{trager1998} \\
		\tioiio & mag & $6066.625$--$6141.625$& $6189.625$--$6272.125$ & $6422.000$--$6455.000$& \citet{la2013} \\
		\naii   & \AA & $8143.000$--$8153.000$&$8180.000$--$8200.000$ & $8233.000$--$8244.000$  & \citet{la2013} \\
		\caii   & \AA & $8474.000$--$8484.000$& $8522.000$--$8562.000$ & $8563.000$--$8577.000$ & \citet{cenarro2001} \\
		\hline
	\end{tabular}
	\label{tab:tab4}
\end{table*}

\subsection{Other indices }

The most useful features for investigating ages in spectra of galaxies are the Hydrogen Balmer lines. Given the spectral range of our data, we employ \hdf \space as an age indicator. Also, we use \mgfep \space index which is adapted for measuring the mean metallicity of galaxies and is independent of the \afe \space \citep{thomas2003}. Moreover, to assess the dependency of \afe \space ratio on the environment  we use \mgb \space and $<$Fe$>$ spectral indices.

\section{RESULTS} \label{sec:results}

In Section~\ref{sec:sigmatrend} we base our analysis entirely on an empirical approach. We first compare the age, metallicity, and abundance sensitive features as a function of \sig in clusters, groups and voids. This needs to be addressed in our study before any assessment of the IMF-sensitive spectral indices to confirm that the possible differences in the IMF indicators of various environments are not simply caused by variations in age, metallicity or abundance ratio. Then we look at the trend of IMF-sensitive indices vs. $\sigma$ for the various samples. In Section~\ref{sec:models} we perform a more quantitative analysis by comparing our measurements with the MILES stellar population models \citep{vazdekis2015}. Note that we do not aim at obtaining a precise determination of IMF-slope, but rather to quantify the differences of IMF-slope ($\Delta\Gamma_{b}$) among low-, intermediate- and high-density environments.

\subsection{Comparing the trend of spectral indices with velocity dispersion} \label{sec:sigmatrend}

\begin{figure}
	
	\includegraphics[width=\linewidth, clip]{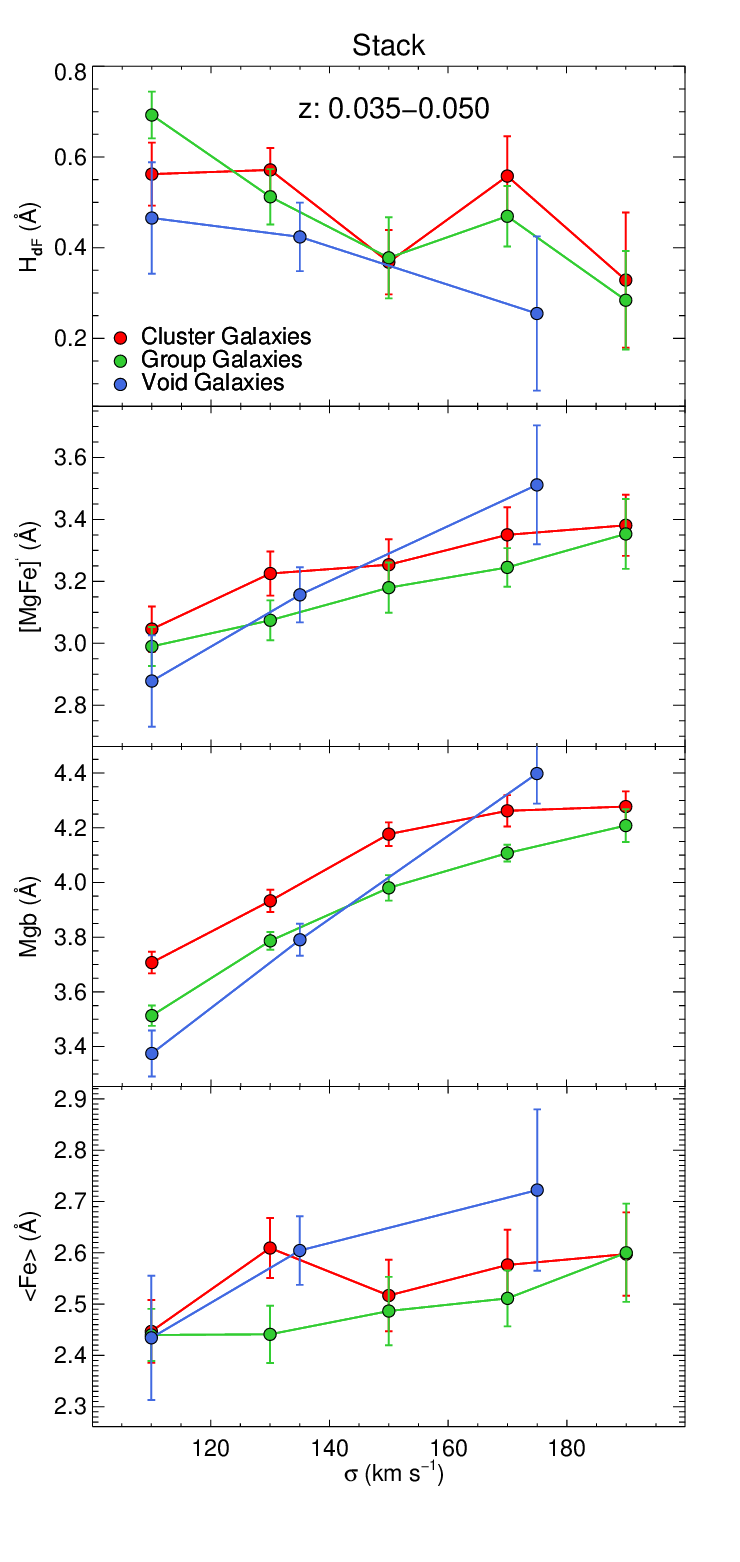}
	\caption{The trend of age, metallicity and [Mg/Fe] indicators with respect to velocity dispersion for the redshift-bin 0.035-0.050. The plots show the index measurements for stacked spectra at each $\sigma$-bin. The red, green and blue colors correspond to cluster, group and void galaxies, respectively. All measurements are performed on data convolved to a common velocity dispersion of 214\kms. The error bars are determined through boot-strapping analysis (see the text for details).}
	\label{fig:fig3}
\end{figure}

Figure~\ref{fig:fig3} shows the measurements for the age-, metallicity- and abundance-sensitive absorption features for the redshift-bin 0.035-0.050. Note that this bin has the best statistics for all of the environments. The results and discussion for the other redshift-bins are shown in Appendix~\ref{app:app2}. This figure shows the index measurements for the stacked spectra of galaxies in different environments. Red, green and blue colors represent the cluster, group and void galaxies, respectively.

The upper panel of Figure~\ref{fig:fig3} compares the variation of the \hdf, which is a proxy of age, with velocity dispersion for galaxies, residing in low, intermediate and high dense environments. In all environments, we see that with increasing $\sigma$, the \hdf \space absorption decreases (about 0.2\AA \space in clusters and voids and about 0.4\AA \space in groups) which is consistent with more massive galaxies being older. Within the error bars, this behavior does not depend on the galactic environment. The result has been checked by determining the median of line-strength of individual galaxies in each bin instead of measuring the line-strength on the stacked spectra (see Appendix~\ref{app:app1}). We find a $\Delta$\hdf$\sim$0.2\AA \space among various environments in the lowest $\sigma$-bin. Assuming solar metallicity, this translates to a $\Delta$age$\sim$1.5Gyr (according to E-MILES SSPs \citep{vazdekis2016}). This small difference in age has a negligible impact on the IMF-sensitive indices and introduces an insignificant uncertainty in the IMF slope ($\Delta\Gamma_{b}\sim0.3$).  Therefore, in our samples, the age difference between galaxies residing in low and high density environments do not cause a notable difference in the strength of IMF-sensitive features among various environments.

In Figure~\ref{fig:fig3}, second panel from the top, probes the environmental dependency on metallicity. In this panel, \mgfep \space is used as a total metallicity indicator. This index rises for all three galaxy samples with increasing velocity dispersion, which means an increase in metallicity. Within the error bars, the various environments behave the same as a function of velocity dispersion. This insensitivity to the environment is more clear in the median panel, where the error bars are smaller (see Appendix~\ref{app:app1}).  However, we see that within the error bars the differences in the metallicity indicator for the three galaxy samples are negligible. For instance, in the [100-120] \kms \space bin, $\Delta$\mgfep \space among various environments is about 0.17\AA. This difference does not affect significantly our IMF indicators and in consequence our IMF slope estimates ($\Delta\Gamma_{b}\sim0.3$).

The behavior of abundance line indicators with respect to \sig is shown in the first and second panels from the bottom of Figure~\ref{fig:fig3}. There is a hint towards a lower [Mg/Fe] in voids for the lowest mass-bins, i.e. [100-120] and [120-140] \kms. But for the other bins the environmental dependency cannot be seen within their respective error bars. This means that the star formation time-scale of lower mass galaxies in dense regions is slightly shorter than in low-density environments.

We have shown in Figure~\ref{fig:fig3} that our index measurements do not suggest any significant age, metallicity or [Mg/Fe] difference among the three galaxy samples, which could introduce a noticeable differential effect on the selected IMF-sensitive indicators. Therefore any differences  between the strengths of IMF-sensitive features in clusters, groups and voids, cannot be attributed to the age, metallicity or [Mg/Fe] abundance differences. We emphasize that this inference is only for the intermediate-mass galaxy regime. 

\begin{figure}
	
	\includegraphics[width=\linewidth, clip]{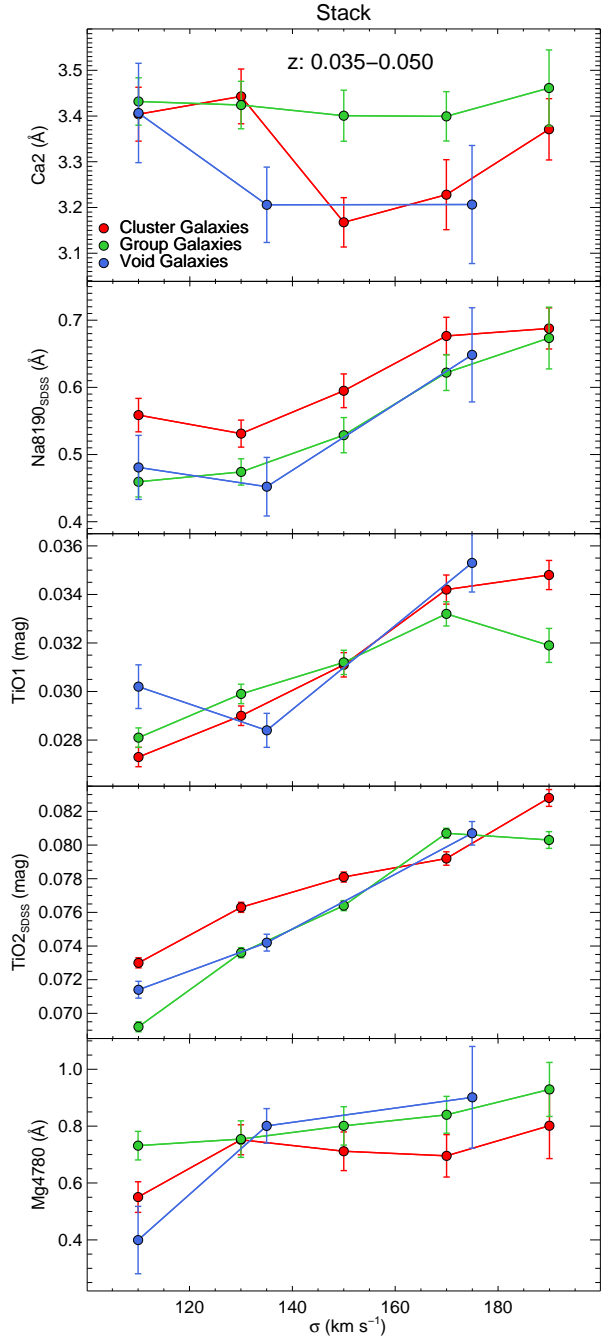}
	\caption{Same as Figure~\ref{fig:fig3}, but for IMF-sensitive line indices.}
	\label{fig:fig4}
\end{figure}

These results put us in a position where we can analyze the spectral indices that are sensitive to the IMF. The behavior of these indices with respect to \sig has been shown in Figure~\ref{fig:fig4}. These plots show the index measurements from the stacked spectra. The first plot from the top shows the absorption line, Ca2, which is mostly sensitive to giant stars. The large scatter seen in this plot can be attributed in part to some residuals of the spectra by telluric absorption and sky emission. We found an average decrease of 0.14 \AA \space for this feature, from \sig= 100 \kms \space to \sig= 200 \kms. This trend is in line with the previous study by \citet{cenarro2003}. These authors found an anti-correlation between \ion{Ca}{ii} triplet indices and log\sig and interpreted this relation as an increase of the dwarf to giant stars ratio along the mass sequence of elliptical galaxies. Also, this amount of average decrease is comparable with the E-MILES SSP models with solar metallicity and varying bimodal IMF slope from 1 to 2.5. In the first plot we see that within the errorbars, the strength of Ca2 in the first $\sigma$-bin is similar for cluster, group and void galaxies. Also, in the last $\sigma$-bin, the measurements for the cluster and group galaxies are similar within the error bar. But in other bins, there are differences in the value of Ca2 at fixed $\sigma$. For instance, in the second $\sigma$-bin, Ca2 is larger in cluster and group galaxies than in voids. However, this behaviour is not confirmed by the median of individual galaxies with the only exception of the third $\sigma$-bin (see Appendix~\ref{app:app1}). The strength of Ca2 in this bin is higher in groups than clusters, both for stack and median panels. However, on the basis of these results we do not find any clear environmental dependency for Ca2. Thus for the second and forth $\sigma$-bins of the left panel, we cannot interpret these differences as an effect of environment on the IMF (see Section~\ref{sec:models}).  

The behavior of the dwarf-sensitive index, $\rm Na8190_{SDSS}$, with respect to \sig is shown in the second panel of Figure~\ref{fig:fig4}. The strength of this absorption line increases with \sig suggesting that galaxies become more dwarf-rich towards higher velocity dispersions. This is consistent with \citet{la2013} who found that there is a trend between IMF slope and $\sigma$, evolving from Kroupa/Chabrier IMF at $\sigma$=100\kms \space to a bottom-heavier IMF with increasing $\sigma$. Their result is based on a large sample of ETGs from the SDSS-based SPIDER survey. Figure 12 of \citet{la2013} shows that the slope of a bimodal IMF changes from 1 to 2.5 for galaxies with velocity dispersions between  100 to 200\kms (\sig range of our sample). Moreover, in their work, the sensitivity of some spectral indices on age, metallicity and IMF slope has been explored with the aid of the extended MILES SSPs. In our galaxy samples this index increases by $\sim0.15$\AA \space from the lowest mass stack to the most massive one. This variation represents an increase of the (bimodal) IMF slope from 1 to 2.5, according to Fig. 4 in \citet{la2013}, which is in good agreement with their work. With the exception of our highest velocity dispersion bin, we found that our cluster galaxy stacks sample shows a slightly larger \naii index value in comparison to the other two environments. However, within the errorbars, this difference ($\sim0.07$\AA \space in average) is not confirmed by the medians of the individual galaxies in each mass bin.

The third panel of Figure~\ref{fig:fig4} displays the variation of \tioi with $\sigma$. The strength of this index increases with increasing  \sig between the two extreme $\sigma$ bins by $\sim0.0048$. This change represents a similar IMF variation as in the case of the $\rm Na8190_{SDSS}$, in good agreement with the study by \citet{la2013}. In the first $\sigma$-bin of stacked spectra, it can be seen that the absorption of \tioi is larger in the void sample than in groups and clusters.  Also, the last $\sigma$-bin shows that the strength of \tioi in cluster galaxies is larger than in groups. But these differences are not confirmed by the median panel (see Appendix~\ref{app:app1}). Thus, in general, the \tioi index behaves similarly as a function of galactic environment.

The \tioiio panel shows that the strength of this line rises about 0.01 mag within our \sig range which is consistent with the variation of IMF slope from 1 to 2.5, in good agreement with the variation resulting from the other two IMF-sensitive indices. From these measurements, it seems that galaxies in cluster regions have larger values of \tioiio than galaxies in groups (except for the $\sigma$-bin of [160-180] \kms). But this is not confirmed by the median plot in Appendix~\ref{app:app1}. Therefore these differences are not meaningful (see Sec.\ref{sec:models} for discussion of differences). Overall the major uncertainties affecting our index measurements include the Poisson noise and the scattering in the index measurements along $\sigma$.

The \mgf index (last panel in Figure~\ref{fig:fig4}) increases slightly with increasing \sig (about 0.2\AA \space in clusters and groups which is again in line with varying IMF slope from 1 to 2.5 in the study by \citet{la2013}). We do not find any significant dependence of this index with the environment except for the lowest-mass bin ($\sigma$=100-120\kms) although this extent is not confirmed in the corresponding median plot in Appendix~\ref{app:app1}. 

Qualitatively, the behaviour of the IMF-sensitive indices among different environments does not show any significant dependence of the IMF-slope on the galactic environment. 

Note that most of the discrepancies between the environments are seen in the lowest $\sigma$-bin. This could be due to the newly accreted low-mass galaxies to the groups and clusters. They might have different accretion histories which makes such an scatter in the index measurements.

\subsection{Comparison of data to model grids} \label{sec:models}

We now give a more quantitative expression for the variation of the spectral indices with galactic environments by comparing the results in the most massive bin - where the maximum change in IMF-slope is likely to be seen - with expectations of models.

We use the MILES stellar population models, covering the spectral range from 3540.5 to 7409.6 \AA, at resolution $\sigma$ = 214 \kms. The MILES SSPs cover a wide range of ages, from 0.03 to 14.00 Gyr, and 12 metallicity bins, i.e. [M/H] = \{-1.79, -1.49, -1.26, -0.96, -0.66, -0.35, -0.25, +0.06, +0.15, +0.26\} with varying \afe (i.e. \afe \space = \{0.00, 0.40\}). The SSPs are provided for several forms of IMFs. We use here the bimodal IMF \citep{vazdekis1996} which is charactrized by its slope as a single free parameter.
	
\begin{figure}
	\includegraphics[width=\linewidth, clip]{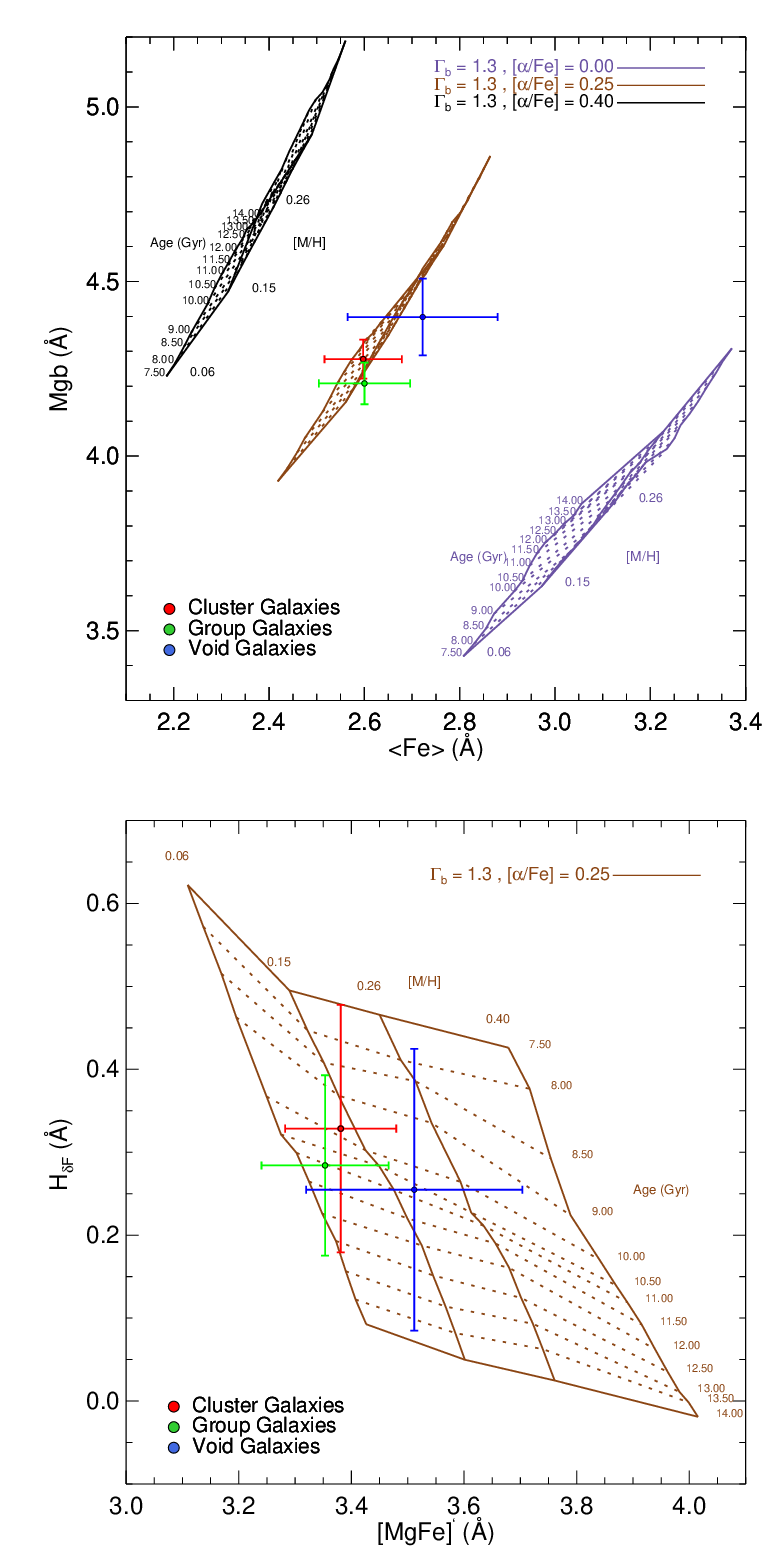}
	\caption{ \textbf{Top} panel shows the \mgb-<Fe> index-index diagram showing positions of the three stacked spectra of the highest $\sigma$-bin with redshifts between 0.035 and 0.050 in our samples (\sig= [180-200] \kms\space for cluster and group galaxies and \sig= [150-200] \kms\space for void galaxies). Over plotted are the MILES SSP models for a Kroupa IMF (both indices are nearly insensitive to IMF). The data points and models are both at a common velocity dispersion of $\sigma \space= 214$ \kms, corresponding to the LIS-8.4 \AA \space resolution. The plot shows that all galaxies are $\alpha$-enhanced with \afe\space $\sim$ 0.25. \textbf{Bottom} panel shows the \hdf-\mgfep\space index-index diagram. The MILES models show variation in age and total metallicity for Kroupa IMF (both indices are again nearly insensitive to IMF). \hdf \space traces variations in age and the \afe-independent \mgfep\space index traces total metallicity. The average age of galaxies in various environments is 10.5 Gyr.}
	\label{fig:fig5}
	
\end{figure}

The upper panel of Figure~\ref{fig:fig5} plots the Mgb vs. <Fe>. All model grids in the Figure~\ref{fig:fig5} refer to the same IMF slope of 1.3 (which resembles closely the Kroupa IMF). The black grid corresponds to the models with \afe \space  = 0.00 and the violet one corresponds to the \afe \space = 0.40. We put the line strengths of the highest velocity dispersion bin of each environment at redshift-bin 0.035-0.050 (\sig= [180-200] \kms\space for cluster and group galaxies and \sig= [150-200] \kms\space for void galaxies); since this bin is the one which we expect the maximum IMF-variation to be seen. Obviously, this plot shows that our sample is $\alpha$-enhanced. Therefore, we linearly interpolated the model grids and found that our measurements match well with the grid of \afe \space = 0.25 (the brown one). So for the rest of our analysis we use the model grid of \afe \space = 0.25. 

\begin{figure}
	\includegraphics[width=\linewidth, clip]{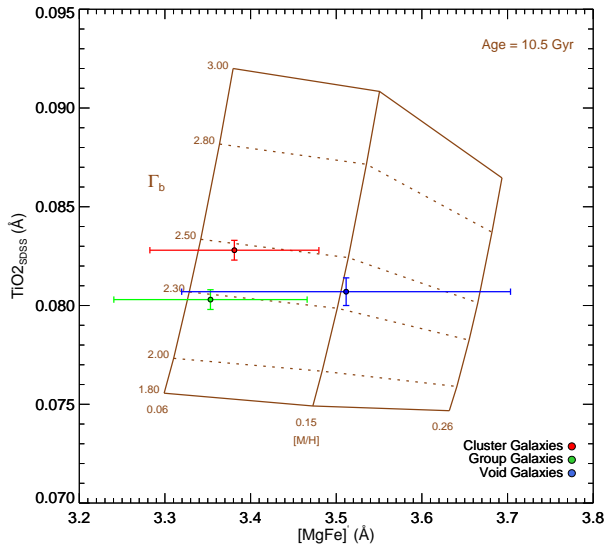}
	\caption{The $\rm TiO_{2_{SDSS}}$-\mgfep \space index-index diagram showing positions of the three stacked spectra of Figure~\ref{fig:fig5}, compared with the stellar population synthesis model predictions of MILES for an age of 10.5 Gyr. This plot shows the difference in IMF-slope for galaxies residing in high-, intermediate-, and low-density environments is  $\Delta\Gamma_{b}\lesssim0.2$.  }
	\label{fig:fig6}
	
\end{figure}

The lower panel of Figure~\ref{fig:fig5} plots the age and metallicity indicators, \hdf \space and \mgfep \space of both the interpolated models (\afe \space = 0.25) and the highest $\sigma$-bin of the cluster, group and void galaxies (in red, green and blue colors, respectively). Here we provide a quantitative measurement of the uncertainty in the derived IMF-slope that takes into account the differences in age and metallicity. Although there is substantial uncertainty in the age of stacked spectra, the average ages of galaxies in three environments is $\sim$ 10.5 Gyr. Thus we make a grid of models with age of 10.5 Gyr in Figure~\ref{fig:fig6} (we also checked that making grids with different ages, within the error bars, does not change significantly the final inferred differences in IMF-slope $\Delta\Gamma_{b}$). In this figure the IMF-sensitive indicator, $\rm TiO_{2_{SDSS}}$, which is insensitive to metallicity (in the metallicity regime characteristic of this galaxy stacks) \citep{la2013} is plotted vs. \mgfep. Although the error bars in \mgfep \space are large (making a very large uncertainty in metallicity estimates), the quantitative amount of inferred IMF variation is negligible among our samples. According to Figure~\ref{fig:fig6}, the largest difference in IMF-slope for galaxies residing in the cluster and group environments is  $\Delta\Gamma_{b_{Cluster-Group}}\sim0.2$,  cluster and void regions $\Delta\Gamma_{b_{Cluster-Void}}<0.2$ and finally void and group one is $\Delta\Gamma_{b_{Void-Group}}\sim0.1$. Hence within this uncertainty we do not find any significant environmental dependency for the stellar IMF. This is consistent with our finding in Sec.\ref{sec:sigmatrend}, inferred by an empirical approach.

Note that this approach is only meant to test the robustness of our finding in Section~\ref{sec:sigmatrend} against uncertainties on line-strength measurements and consequently age/metallicity of the samples.

\section{SUMMARY AND DISCUSSION} \label{sec:discussion}

In this work we have studied the possible influence of large-scale environment on the stellar IMF of galaxies. For this purpose, we have used a sample of local quiescent galaxies from \citet{mosleh2018}, which consists of three sub-samples of isolated quiescent void galaxies and quiescent group/cluster galaxies with the same stellar mass distribution as of the voids sample. We separated the galaxies in each sub-sample according to different \sig and redshift bins and stacked their SDSS DR7 spectra to achieve high signal-to-noise ratio, which is required to analyze the IMF-sensitive spectral indices. Our covered \sig range ($100<$ \sig $<200$ \kms) is imposed by our void galaxy sample due to the fact that these regions do not host galaxies as massive as those in groups/clusters. Therefore, our analysis of IMF sensitive features has been limited to the intermediate-mass galaxies.  

We measured various line strength indices and compared their trends with respect to $\sigma$, among three different environments. As a first step, we investigated the trends of age/metallicity line-sensitive indices that constrain these relevant stellar population properties. We do not find environmental dependences in our age-, metallicity- and [Mg/Fe]-sensitive indices that might translate to non negligible variations in our IMF-sensitive indices. Hence, any possible dependence of the IMF-sensitive features on environment cannot be fully attributed to the variation of these stellar population parameters. These results allowed us to study the trends of the selected IMF indicators as a function of velocity dispersion for our three galaxy samples. We find that for the covered \sig range of intermediate stellar mass systems, these trends do not show any significant dependence on the large-scale environment of galaxies, within our index measurements precision. To provide a more quantitative estimate of the uncertainties, we compared our results for the highest $\sigma$-bin of each environment to the predictions of MILES SSP models. The obtained results suggest that the IMF is not affected significantly by the environment where our intermediate-mass galaxies reside. This result is in line with the recent work by \citet{rosani2018}, despite the fact that their definition of environment is different than ours. In their work, they used SDSS spectra of a sample of ETGs from the SPIDER catalog and made use of state-of-the-art stellar population synthesis models to constrain quantitatively the IMF slope. They used host halo mass as a proxy for the environment and found that there is no dependence of the IMF slope on the environment.

Further studies, with more sensitive data and different methods, involving more massive objects, should be performed to further constrain the role of environment on the IMF.

\section*{ACKMOWLEDGEMENTS}

We thank Alireza Molaeinezhad for the help with stacking spectra and also Jesus Falc{\'o}n-Barroso for providing us his pixel-by-pixel smoothing code. We also thank the anonymous referee for a constructive report that helped us to improve our manuscript. EE and AV acknowledge support from grant AYA2016-77237-C3-1-P from the Spanish Ministry of Economy and Competitiveness (MINECO). This paper is based on data retrieved from the Sloan Digital Sky Survey archives (\url{http://classic.sdss.org/collaboration/credits.html}). Funding for the SDSS and SDSS-II has been provided by the Alfred P. Sloan Foundation, the Participating Institutions, the National Science Foundation, the U.S. Department of Energy, the National Aeronautics and Space Administration, the Japanese Monbukagakusho, the Max Planck Society, and the Higher Education Funding Council for England. 




\bibliographystyle{mnras}
\bibliography{references} 



\appendix

\section{Comparison of measurements on the stacked spectra with estimates on individual galaxies} \label{app:app1}

\begin{figure}
	
	\includegraphics[width=\linewidth, clip]{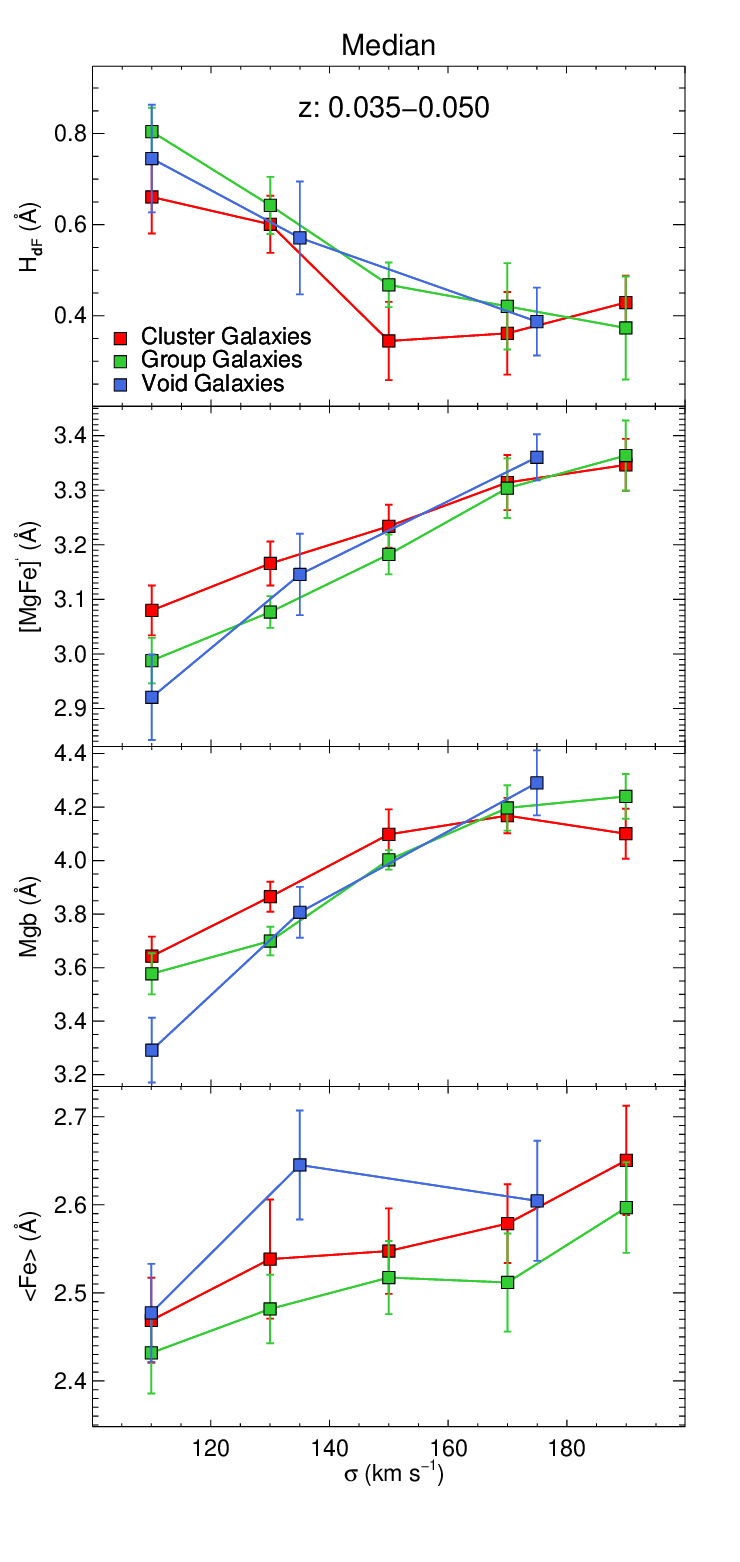}
	\caption{The trend of age, metallicity and [Mg/Fe] indicators with respect to velocity dispersion for the redshift-bin 0.035-0.050. The plots show the medians of individual galaxies in each bin. The red, green and blue colors correspond to cluster, group and void galaxies, respectively. All measurements are performed on data convolved to a common velocity dispersion of 214\kms. The error bars are determined through boot-strapping analysis.}
	\label{fig:fig7}
\end{figure}

\begin{figure}
	
	\includegraphics[width=\linewidth, clip]{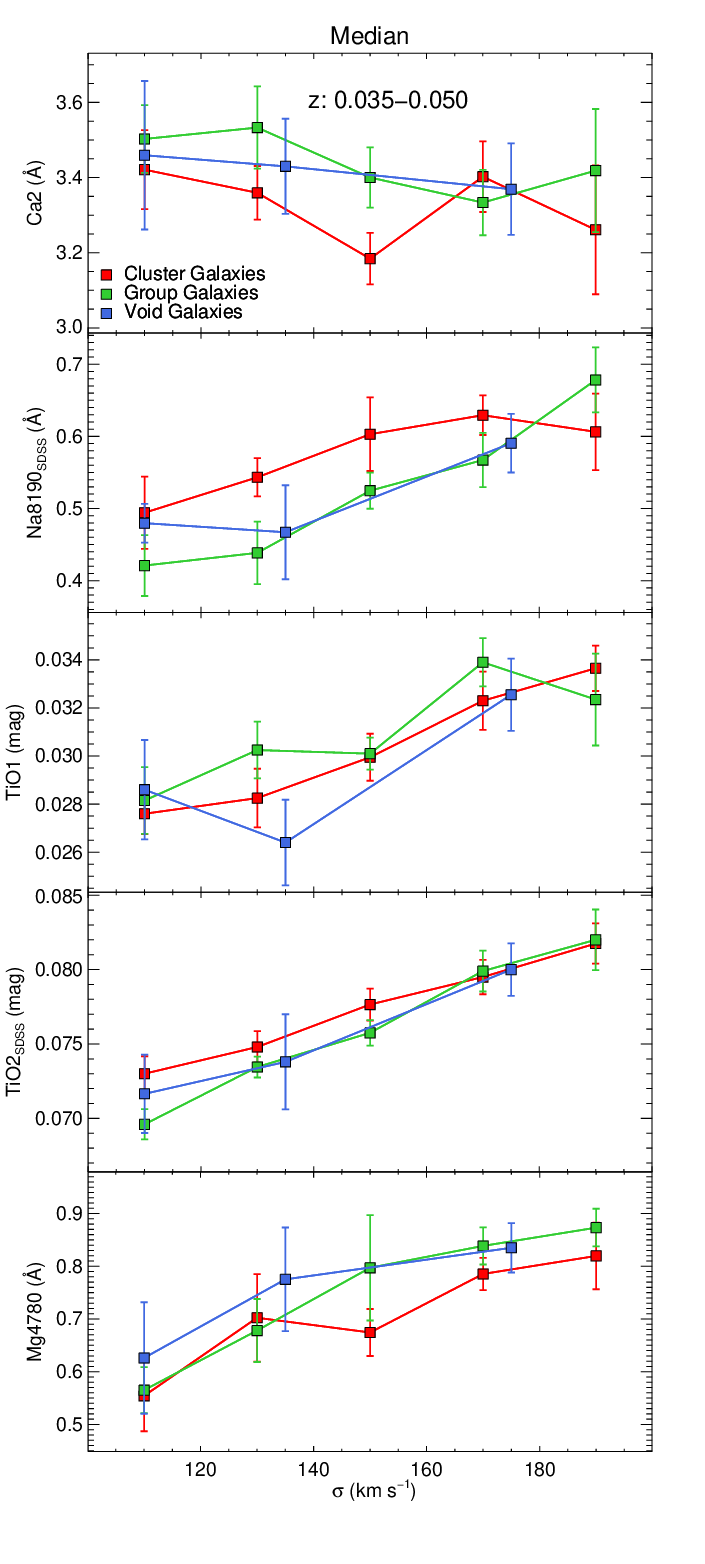}
	\caption{Same as Figure~\ref{fig:fig7}, but for IMF-sensitive line indices.}
	\label{fig:fig8}
\end{figure}

To further investigate the robustness of our stacking procedure, we measure the median of line-strength of individual galaxies in each $\sigma$-bin of Figures \ref{fig:fig3} and \ref{fig:fig4}. in Figures \ref{fig:fig7} and \ref{fig:fig8}, we recover well the trend of spectral indices with $\sigma$. These plots help us to disentangle the environmental differences of spectral features from the systematic ones.  

\section{Other Redshift-Bins} \label{app:app2}

\subsection{Redshift-bin $0.020<z<0.035$}

\begin{figure*}

	\includegraphics[width=\linewidth, clip]{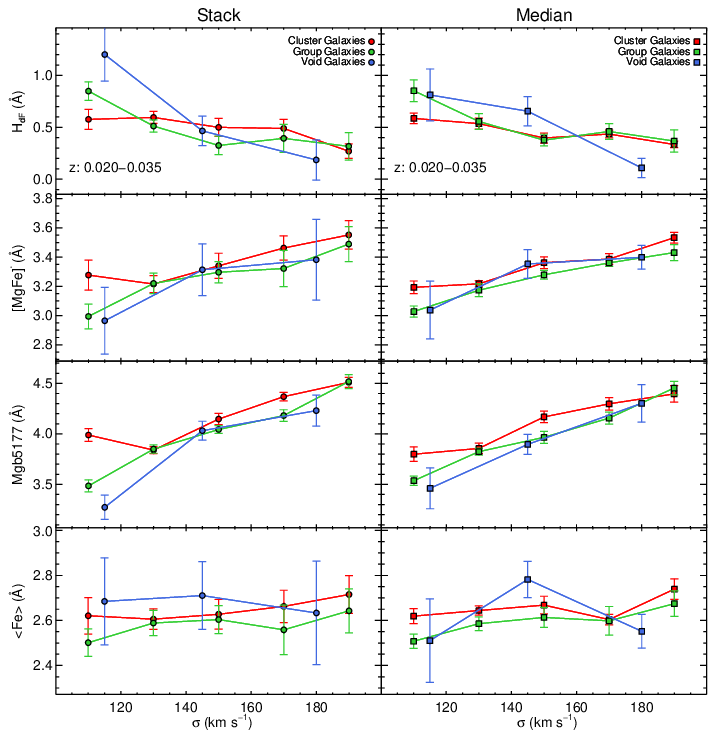}
	\caption{The trend of age, metallicity and [Mg/Fe] indicators with respect to velocity dispersion for the redshift-bin 0.020-0.035. The left panels show the results for the stacked spectra while the right panels show the medians of individual galaxies in each bin. The red, green and blue colors correspond to cluster, group and void galaxies, respectively. All measurements are performed on data convolved to a common velocity dispersion of 214\kms. The error bars are determined through boot-strapping analysis.}
	\label{fig:fig9}
\end{figure*}

We show in Figure~\ref{fig:fig9} the trend of age, metallicity and [Mg/Fe] indicators with \sig for the redshift bin of $0.020<z<0.035$. The left column shows the index measurements from the stacked spectra and the right column shows the median of individual measurements  for testing the robustness of the stacking procedure. We see differences between index values in various environments, particularly for the lowest $\sigma$-bin ([100-120] \kms). Note however that these differences are within the error bars. Therefore, possible variations of IMF-sensitive indices in different environments cannot be fully attributed to the variations of age, metallicity or [Mg/Fe].

\begin{figure*}

	\includegraphics[width=\linewidth, clip]{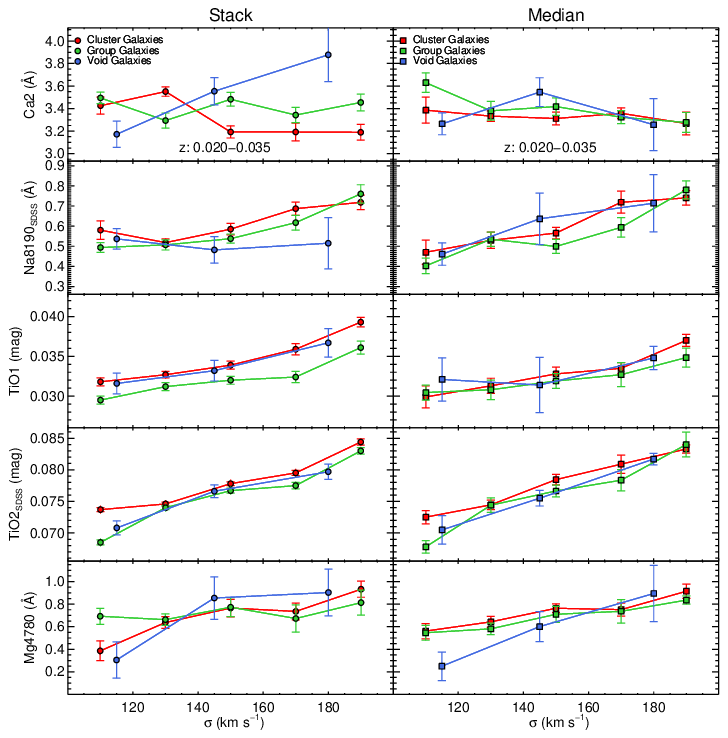}	
	\caption{Same as Figure~\ref{fig:fig9}, but for IMF-sensitive line indices.}
	\label{fig:fig10}		
\end{figure*}

The behaviour of IMF-sensitive features with $\sigma$, is shown in Figure~\ref{fig:fig10} for the redshift bin $0.020<z<0.035$. The first panel from the top corresponds to the Ca2 absorption line. In the panel showing the measurements for the stacked spectra, the strength of this index decreases with increasing \sig in cluster and group galaxies in agreement with \citet{cenarro2003}. However we find that in voids it increases, although this extent is not confirmed in the right panel showing the medians. These contradicting results can be fully attributed to poor statistics (only 19 galaxies in total) of our void sample in this redshift bin.

Regarding $\rm TiO_{2_{SDSS}}$, we find a strong increase with \sig in all galactic environments but there is a noticeable difference between cluster and group galaxies in the lowest $\sigma$-bin. The median panel shows the same discrepancy. Half of the difference in the strength of \tioiio could be attributed to an age difference ($\sim$2Gyr) between the cluster and group samples in this bin. Note that in this bin there is a difference in the index value of \mgfep \space among cluster and group galaxies, which means a difference in metallicty among these two samples. However, the observed difference in \tioiio could not be attributed to the metallicity difference; because \tioiio is completely insensitive to metallicty. 

Within our precision we do not find a robust indication of a dependence of our IMF-sensitive indices.

\subsection{Redshift-bin $0.050<z<0.055$}

\begin{figure*}

	\includegraphics[width=\linewidth, clip]{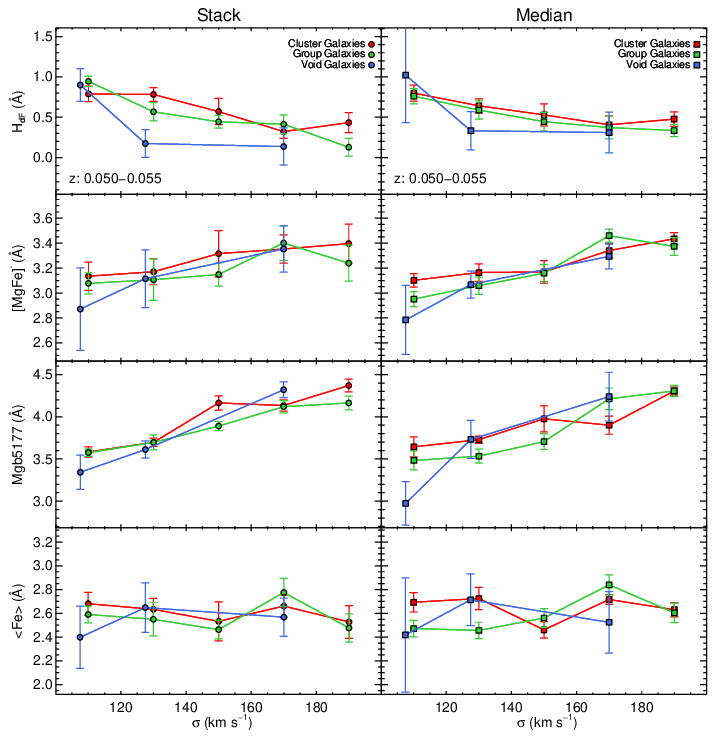}
	\caption{The trend of age, metallicity and [Mg/Fe] indicators with respect to velocity dispersion for the redshift-bin 0.050-0.055. The left panels show the results for the stacked spectra while the right panels show the medians of individual galaxies in each bin. The red, green and blue colors correspond to cluster, group and void galaxies, respectively. All measurements are performed on data convolved to a common velocity dispersion of 214\kms. The error bars are determined through boot-strapping analysis.}
	\label{fig:fig11}
\end{figure*}

Figure~\ref{fig:fig11} shows the behavior of indicators of stellar population parameters as a function of \sig for galaxies within the redshift bin 0.050-0.055. We see minor differences in the values of \hdf, \mgfep, \mgb \space and $<$Fe$>$ among the various galactic environments. Within their respective error bars, no trend with the environment is observed.

\begin{figure*}
	\includegraphics[width=\linewidth, clip]{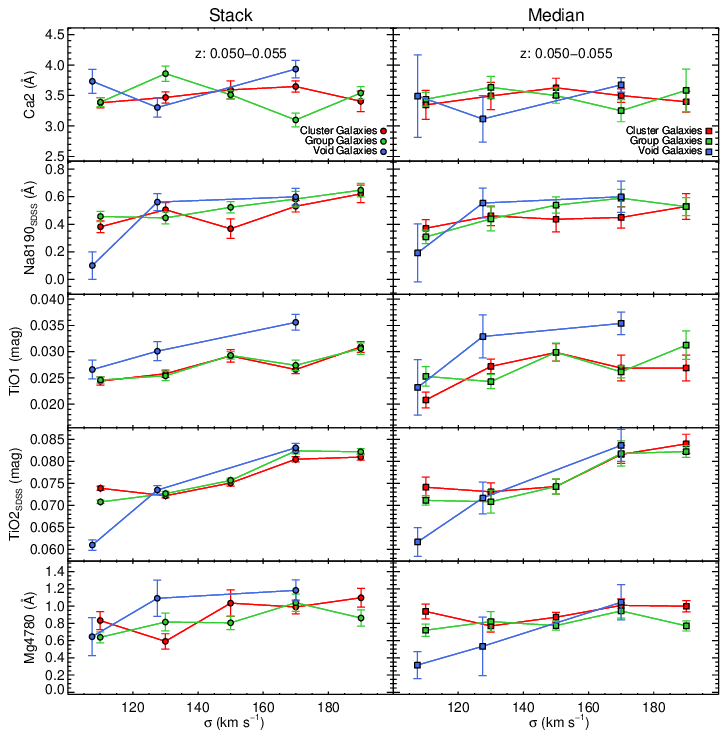}
	\caption{Same as Figure~\ref{fig:fig11}, but for IMF-sensitive line indices.}
	\label{fig:fig12}
\end{figure*}

The last Figure (Figure~\ref{fig:fig10}), shows variations of IMF-sensitive absorption lines with $\sigma$. The first panel from the top corresponds to the Ca2 line. A large scatter can be seen in the trend of this line with $\sigma$. However within the error bars and compared to the panel of medians, the strength of this index is similar in cluster, group and void galaxies. 

There is a large discrepancy in \tioi between isolated void galaxies and the other samples. This apparent difference is likely due to the low number of galaxies in this redshift bin compared to the group and cluster samples. 

Also in the first $\sigma$-bin of \tioiio we find a significant difference among void galaxies and other two samples. This can be totally attributed to the poor statistics of the void sample. 


\section{Line-strength measurements} \label{app:app3}

We report measurements of stellar population parameter indicators for our stacked spectra in Tables \ref{tab:tab5}, \ref{tab:tab6} and \ref{tab:tab7}. Also, in Tables \ref{tab:tab8}, \ref{tab:tab9} and \ref{tab:tab10}, we present the line-strength of IMF-sensitive indices measured from the stacked spectra of galaxies in clusters, groups and voids, respectively. The indices are measure in the stacked spectra at $\sigma=214$\kms. 

\begin{table*}
	\caption{Measurements of the age, metallicity and [Mg/Fe] indicators for the stacked spectra of cluster galaxies in bins of redshift and velocity dispersion}
	\begin{tabular}{c c c c c c}
		\hline
		
		redshift  & \sig & \hdf & \mgfep & \mgb & $<$Fe$>$  \\
		& $\rm [km/s]$ & [\AA] & [\AA] & [\AA] & [\AA]  \\
		\hline
		\multirow{5}{*}{$0.020$--$0.035$}&	$100$--$120$ &0.58 $\pm$     0.10 &       3.28 $\pm$      0.10 &
		3.99 $\pm$     0.06 &       2.62 $\pm$     0.08 \\
		
		&	$120$--$140$&  0.60 $\pm$     0.06 &       3.21 $\pm$     0.06 &
		3.84 $\pm$     0.04 &       2.60 $\pm$     0.04 \\

		&	$140$--$160$& 0.50 $\pm$     0.09 &       3.34 $\pm$     0.09 &
		4.15 $\pm$     0.05 &       2.63 $\pm$     0.06 \\
		
		&	$160$--$180$&  0.49 $\pm$     0.09 &       3.46 $\pm$     0.08 &
		4.37 $\pm$     0.04 &       2.66 $\pm$     0.07 \\

		&	$180$--$200$&   0.27 $\pm$     0.07 &       3.55 $\pm$     0.10 &
		4.51 $\pm$     0.05 &       2.71 $\pm$     0.08 \\

		\hline
		\multirow{5}{*}{$0.035$--$0.050$}&	$100$--$120$ &   0.56 $\pm$     0.07 &       3.05 $\pm$     0.07 &
		3.70 $\pm$     0.04 &       2.45 $\pm$     0.06 \\

		&	$120$--$140$& 0.57 $\pm$     0.05 &       3.22 $\pm$     0.07 &
		3.93 $\pm$     0.04 &       2.61 $\pm$     0.06 \\

		&	$140$--$160$& 0.37 $\pm$     0.07 &       3.25 $\pm$     0.08 &
		4.18 $\pm$     0.04 &       2.52 $\pm$     0.07 \\
		
		&	$160$--$180$&  0.56 $\pm$     0.09 &       3.35 $\pm$     0.09 &
		4.26 $\pm$     0.06 &       2.58 $\pm$     0.07 \\

		&	$180$--$200$&   0.33 $\pm$      0.15 &       3.38 $\pm$     0.10 &
		4.28 $\pm$     0.06 &       2.60 $\pm$     0.08 \\

		\hline
		\multirow{5}{*}{$0.050$--$0.055$}&	$100$--$120$&  0.79 $\pm$     0.09 &       3.13 $\pm$      0.11 &
		3.58 $\pm$     0.06 &       2.68 $\pm$     0.09 \\
		
		&	$120$--$140$&  0.78 $\pm$     0.09 &       3.17 $\pm$      0.10 &
		3.70 $\pm$     0.05 &       2.63 $\pm$     0.09 \\

		&	$140$--$160$& 0.57 $\pm$      0.16 &       3.32 $\pm$      0.19 &
		4.16 $\pm$     0.09 &       2.53 $\pm$      0.16 \\
		
		&	$160$--$180$&   0.32 $\pm$     0.08 &       3.35 $\pm$      0.11 &
		4.13 $\pm$     0.06 &       2.66 $\pm$     0.09 \\

		&	$180$--$200$&   0.43 $\pm$      0.12 &       3.40 $\pm$      0.16 &
		4.37 $\pm$     0.07 &       2.53 $\pm$      0.14 \\

		\hline
	\end{tabular}
	\label{tab:tab5}
\end{table*}

\begin{table*}
	\caption{Measurements of the age, metallicity and [Mg/Fe] indicators for the stacked spectra of group galaxies in bins of redshift and velocity dispersion}
	\begin{tabular}{c c c c c c}
		\hline
		
		redshift  & \sig & \hdf & \mgfep & \mgb & $<$Fe$>$  \\
		& $\rm [km/s]$ & [\AA] & [\AA] & [\AA] & [\AA]  \\
		\hline
		\multirow{5}{*}{$0.020$--$0.035$}&	$100$--$120$ &0.85 $\pm$     0.09 &       2.99 $\pm$     0.08 &
		3.49 $\pm$     0.06 &       2.50 $\pm$     0.06 \\
		
		&	$120$--$140$&  0.51 $\pm$     0.06 &       3.22 $\pm$     0.07 &
		3.85 $\pm$     0.04 &       2.59 $\pm$     0.06 \\

		&	$140$--$160$& 0.33 $\pm$     0.09 &       3.30 $\pm$     0.07 &
		4.04 $\pm$     0.04 &       2.60 $\pm$     0.06 \\
		
		&	$160$--$180$&   0.39 $\pm$      0.13 &       3.32 $\pm$      0.12 &
		4.18 $\pm$     0.06 &       2.56 $\pm$      0.11 \\

		&	$180$--$200$&    0.32 $\pm$      0.13 &       3.49 $\pm$      0.12 &
		4.52 $\pm$     0.07 &       2.64 $\pm$     0.10 \\

		\hline
		\multirow{5}{*}{$0.035$--$0.050$}&	$100$--$120$ &   0.70 $\pm$     0.05 &       2.99 $\pm$     0.06 &
		3.51 $\pm$     0.04 &       2.44 $\pm$     0.05 \\

		&	$120$--$140$&  0.51 $\pm$     0.06 &       3.07 $\pm$     0.06 &
		3.79 $\pm$     0.03 &       2.44 $\pm$     0.05 \\

		&	$140$--$160$&  0.38 $\pm$     0.09 &       3.18 $\pm$     0.08 &
		3.98 $\pm$     0.05 &       2.49 $\pm$     0.07 \\
		
		&	$160$--$180$& 0.47 $\pm$     0.07 &       3.25 $\pm$     0.06 &
		4.10 $\pm$     0.03 &       2.51 $\pm$     0.05 \\

		&	$180$--$200$&   0.28 $\pm$      0.11 &       3.35 $\pm$      0.11 &
		4.21 $\pm$     0.06 &       2.60 $\pm$     0.10 \\

		\hline
		\multirow{5}{*}{$0.050$--$0.055$}&	$100$--$120$&   0.94 $\pm$     0.07 &       3.08 $\pm$     0.09 &
		3.57 $\pm$     0.04 &       2.59 $\pm$     0.07 \\

		&	$120$--$140$&  0.57 $\pm$      0.11 &       3.11 $\pm$      0.16 &
		3.70 $\pm$     0.09 &       2.55 $\pm$      0.14 \\

		&	$140$--$160$& 0.44 $\pm$     0.08 &       3.15 $\pm$     0.09 &
		3.89 $\pm$     0.05 &       2.46 $\pm$     0.08 \\
		
		&	$160$--$180$&   0.41 $\pm$      0.12 &       3.40 $\pm$      0.14 &
		4.12 $\pm$     0.07 &       2.77 $\pm$      0.12 \\
		
		&	$180$--$200$&  0.13 $\pm$      0.11 &       3.24 $\pm$      0.14 &
		4.16 $\pm$     0.08 &       2.48 $\pm$      0.12 \\

		\hline
	\end{tabular}
	\label{tab:tab6}
\end{table*}

\begin{table*}
	\caption{Measurements of the age, metallicity and [Mg/Fe] indicators for the stacked spectra of void galaxies in bins of redshift and velocity dispersion}
	\begin{tabular}{c c c c c c}
	\hline
	
	 redshift  & \sig & \hdf & \mgfep & \mgb & $<$Fe$>$  \\
	 & $\rm [km/s]$ & [\AA] & [\AA] & [\AA] & [\AA]  \\
		\hline
		
		\multirow{3}{*}{$0.020$--$0.035$}& $100$--$130$	&      1.20 $\pm$      0.26 &       2.96 $\pm$      0.23 &
		3.27 $\pm$      0.12 &       2.68 $\pm$      0.19 \\

		&	$130$--$160$&  0.47 $\pm$      0.14 &       3.31 $\pm$      0.18 &
		4.03 $\pm$     0.09 &       2.71 $\pm$      0.15 \\

		&	$160$--$200$& 0.19 $\pm$      0.19 &       3.38 $\pm$      0.28 &
		4.23 $\pm$      0.15 &       2.63 $\pm$      0.23 \\

		\hline 
		
		\multirow{3}{*}{$0.035$--$0.050$}& $100$--$120$	&  0.46 $\pm$      0.12 &       2.88 $\pm$      0.15 &
		3.37 $\pm$     0.08 &       2.43 $\pm$      0.12 \\

		&	$120$--$150$&   0.42 $\pm$     0.08 &       3.16 $\pm$     0.09 &
		3.79 $\pm$     0.06 &       2.60 $\pm$     0.07 \\

		&	$150$--$200$&  0.25 $\pm$      0.17 &       3.51 $\pm$      0.19 &
		4.40 $\pm$      0.11 &       2.72 $\pm$      0.16 \\

		\hline 
		
		\multirow{3}{*}{$0.050$--$0.055$}&	$100$--$115$ &     0.90 $\pm$      0.20 &       2.87 $\pm$      0.33 &
		3.34 $\pm$      0.20 &       2.40 $\pm$      0.26 \\

		&	$115$--$140$&  0.18 $\pm$      0.17 &       3.11 $\pm$      0.23 &
		3.61 $\pm$      0.10 &       2.65 $\pm$      0.21 \\
		
		&	$140$--$200$& 0.14 $\pm$      0.23 &       3.35 $\pm$      0.19 &
		4.32 $\pm$     0.09 &       2.57 $\pm$      0.16 \\

		\hline 
		
	\end{tabular}
	\label{tab:tab7}
\end{table*}


\begin{table*}
	\caption{Strength of the IMF-sensitive features for the stacked spectra of cluster galaxies in bins of  redshift and velocity dispersion}
	\begin{tabular}{c c c c c c c}
		\hline
		
		redshift  & \sig & Ca2 & \naii & \tioi & \tioiio & \mgf \\
		& $\rm [km/s]$ & [\AA] & [\AA] & [mag] & [mag] & [\AA]  \\
		\hline
		\multirow{5}{*}{$0.020$--$0.035$}&	$100$--$120$ &3.43 $\pm$     0.07 &      0.58 $\pm$     0.04 &     0.0318 $\pm$   0.0005 &     0.0737 $\pm$
		0.0003 &      0.39 $\pm$     0.09 \\
		
		&	$120$--$140$&  3.55 $\pm$     0.04 &      0.52 $\pm$     0.02 &     0.0327 $\pm$   0.0004 &     0.0746 $\pm$
		0.0003 &      0.64 $\pm$     0.05 \\
		
		&	$140$--$160$&  3.19 $\pm$     0.05 &      0.59 $\pm$     0.03 &     0.0339 $\pm$   0.0005 &     0.0778 $\pm$
		0.0003 &      0.76 $\pm$     0.08 \\
		
		&	$160$--$180$&  3.19 $\pm$     0.08 &      0.69 $\pm$     0.03 &     0.0359 $\pm$   0.0007 &     0.0795 $\pm$
		0.0004 &      0.73 $\pm$     0.07 \\ 
		
		&	$180$--$200$&  3.19 $\pm$     0.07 &      0.72 $\pm$     0.04 &     0.0393 $\pm$   0.0006 &     0.0844 $\pm$
		0.0005 &      0.93 $\pm$     0.07 \\

		\hline
		\multirow{5}{*}{$0.035$--$0.050$}&	$100$--$120$ &3.40 $\pm$     0.06 &      0.56 $\pm$     0.02 &     0.0273 $\pm$   0.0004 &     0.0730 $\pm$
		0.0003 &      0.55 $\pm$     0.05 \\
		
		&	$120$--$140$&   3.44 $\pm$     0.06 &      0.53 $\pm$     0.02 &     0.0290 $\pm$   0.0004 &     0.0763 $\pm$
		0.0003 &      0.75 $\pm$     0.05 \\
		
		&	$140$--$160$&   3.17 $\pm$     0.05 &      0.59 $\pm$     0.03 &     0.0311 $\pm$   0.0005 &     0.0781 $\pm$
		0.0003 &      0.71 $\pm$     0.07 \\
		
		&	$160$--$180$&   3.23 $\pm$     0.08 &      0.68 $\pm$     0.03 &     0.0342 $\pm$   0.0006 &     0.0792 $\pm$
		0.0004 &      0.70 $\pm$     0.07 \\
		
		&	$180$--$200$& 3.37 $\pm$     0.07 &      0.69 $\pm$     0.03 &     0.0348 $\pm$   0.0006 &     0.0828 $\pm$
		0.0005 &      0.80 $\pm$      0.11 \\

		\hline
		\multirow{5}{*}{$0.050$--$0.055$}&	$100$--$120$ &3.38 $\pm$     0.08 &      0.38 $\pm$     0.04 &     0.0244 $\pm$   0.0008 &     0.0739 $\pm$
		0.0005 &      0.83 $\pm$      0.10 \\
		
		&	$120$--$140$& 3.47 $\pm$     0.09 &      0.51 $\pm$     0.05 &     0.0258 $\pm$   0.0007 &     0.0722 $\pm$
		0.0005 &      0.59 $\pm$     0.09 \\
		
		&	$140$--$160$&  3.59 $\pm$      0.15 &      0.37 $\pm$     0.07 &     0.0292 $\pm$    0.0012 &     0.0751 $\pm$
		0.0007 &       1.03 $\pm$      0.15 \\
		
		&	$160$--$180$& 3.65 $\pm$     0.09 &      0.53 $\pm$     0.04 &     0.0266 $\pm$   0.0008 &     0.0805 $\pm$
		0.0006 &      0.99 $\pm$     0.08 \\
		
		&	$180$--$200$&   3.40 $\pm$      0.17 &      0.62 $\pm$     0.06 &     0.0309 $\pm$    0.0010 &     0.0810 $\pm$
		0.0007 &       1.10 $\pm$      0.11 \\

		\hline
	\end{tabular}
	\label{tab:tab8}
\end{table*}

\begin{table*}
	\caption{Strength of the IMF-sensitive features for the stacked spectra of group galaxies in bins of  redshift and velocity dispersion}
	\begin{tabular}{c c c c c c c}
		\hline
		
		redshift  & \sig & Ca2 & \naii & \tioi & \tioiio & \mgf \\
		& $\rm [km/s]$ & [\AA] & [\AA] & [mag] & [mag] & [\AA]  \\
		\hline
		\multirow{5}{*}{$0.020$--$0.035$}&	$100$--$120$ &3.49 $\pm$     0.05 &      0.49 $\pm$     0.02 &     0.0295 $\pm$   0.0005 &     0.0685 $\pm$
		0.0004 &      0.69 $\pm$     0.07 \\
		
		&	$120$--$140$&  3.29 $\pm$     0.07 &      0.51 $\pm$     0.02 &     0.0312 $\pm$   0.0005 &     0.0740 $\pm$
		0.0003 &      0.66 $\pm$     0.05 \\
		
		&	$140$--$160$&  3.48 $\pm$     0.06 &      0.54 $\pm$     0.02 &     0.0320 $\pm$   0.0005 &     0.0767 $\pm$
		0.0004 &      0.77 $\pm$     0.08 \\
		
		&	$160$--$180$&  3.34 $\pm$     0.07 &      0.62 $\pm$     0.04 &     0.0324 $\pm$   0.0007 &     0.0775 $\pm$
		0.0005 &      0.67 $\pm$      0.12 \\ 
		
		&	$180$--$200$&  3.45 $\pm$     0.07 &      0.76 $\pm$     0.05 &     0.0361 $\pm$   0.0008 &     0.0830 $\pm$
		0.0005 &      0.81 $\pm$      0.11 \\

		\hline
		\multirow{5}{*}{$0.035$--$0.050$}&	$100$--$120$ &3.43 $\pm$     0.05 &      0.46 $\pm$     0.02 &     0.0281 $\pm$   0.0004 &     0.0692 $\pm$
		0.0003 &      0.73 $\pm$     0.05 \\
		
		&	$120$--$140$&  3.42 $\pm$     0.05 &      0.47 $\pm$     0.02 &     0.0299 $\pm$   0.0004 &     0.0736 $\pm$
		0.0003 &      0.75 $\pm$     0.06 \\
		
		&	$140$--$160$&   3.40 $\pm$     0.05 &      0.53 $\pm$     0.03 &     0.0312 $\pm$   0.0005 &     0.0764 $\pm$
		0.0003 &      0.80 $\pm$     0.07 \\
		
		&	$160$--$180$&   3.40 $\pm$     0.05 &      0.62 $\pm$     0.03 &     0.0332 $\pm$   0.0005 &     0.0807 $\pm$
		0.0003 &      0.84 $\pm$     0.06 \\
		
		&	$180$--$200$&  3.46 $\pm$     0.08 &      0.67 $\pm$     0.05 &     0.0319 $\pm$   0.0007 &     0.0803 $\pm$
		0.0005 &      0.93 $\pm$     0.09 \\

		\hline
		\multirow{5}{*}{$0.050$--$0.055$}&	$100$--$120$ & 3.39 $\pm$     0.07 &      0.46 $\pm$     0.04 &     0.0246 $\pm$   0.0006 &     0.0708 $\pm$
		0.0004 &      0.64 $\pm$     0.06 \\
		
		&	$120$--$140$&  3.86 $\pm$      0.12 &      0.45 $\pm$     0.04 &     0.0254 $\pm$   0.0009 &     0.0727 $\pm$
		0.0005 &      0.81 $\pm$      0.10 \\

		&	$140$--$160$&  3.51 $\pm$     0.07 &      0.52 $\pm$     0.04 &     0.0293 $\pm$   0.0007 &     0.0757 $\pm$
		0.0005 &      0.80 $\pm$     0.08 \\
		
		&	$160$--$180$&  3.10 $\pm$      0.11 &      0.58 $\pm$     0.05 &     0.0274 $\pm$    0.0010 &     0.0824 $\pm$
		0.0007 &       1.04 $\pm$     0.10 \\
		
		&	$180$--$200$&   3.54 $\pm$      0.10 &      0.65 $\pm$     0.05 &     0.0306 $\pm$    0.0011 &     0.0822 $\pm$
		0.0007 &      0.86 $\pm$     0.10 \\

		\hline
	\end{tabular}
	\label{tab:tab9}
\end{table*}

\begin{table*}
	\caption{Strength of the IMF-sensitive features for the stacked spectra of void galaxies in bins of redshift and velocity dispersion}
	\begin{tabular}{c c c c c c c}
		\hline
		redshift & \sig & Ca2 & \naii & \tioi & \tioiio & \mgf \\
		& $\rm [km/s]$   &[\AA]& [\AA] & [mag] & [mag]   & [\AA] \\
		\hline
		
		\multirow{3}{*}{$0.020$--$0.035$}& $100$--$130$	&  3.17 $\pm$      0.12 &      0.54 $\pm$     0.05 &     0.0316 $\pm$    0.0013 &     0.0708 $\pm$
		0.0011 &      0.30 $\pm$      0.16 \\
		
		&	$130$--$160$&  3.55 $\pm$      0.12 &      0.48 $\pm$     0.06 &     0.0332 $\pm$    0.0013 &     0.0766 $\pm$
		0.0010 &      0.85 $\pm$      0.19 \\
		
		&	$160$--$200$& 3.88 $\pm$      0.24 &      0.51 $\pm$      0.13 &     0.0367 $\pm$    0.0018 &     0.0797 $\pm$
		0.0012 &      0.90 $\pm$      0.21 \\

		\hline 
		
		\multirow{3}{*}{$0.035$--$0.050$}& $100$--$120$	&   3.41 $\pm$      0.11 &      0.48 $\pm$     0.05 &     0.0302 $\pm$   0.0009 &     0.0714 $\pm$
		0.0005 &      0.40 $\pm$      0.12 \\

		&	$120$--$150$&  3.21 $\pm$     0.08 &      0.45 $\pm$     0.04 &     0.0284 $\pm$   0.0007 &     0.0742 $\pm$
		0.0005 &      0.80 $\pm$     0.06 \\
		
		&	$150$--$200$&   3.20 $\pm$      0.13 &      0.65 $\pm$     0.07 &     0.0353 $\pm$    0.0012 &     0.0807 $\pm$
		0.0007 &      0.90 $\pm$      0.18 \\

		\hline 
		
		\multirow{3}{*}{$0.050$--$0.055$}&	$100$--$115$ &  3.73 $\pm$      0.20 &      0.10 $\pm$      0.10 &     0.0266 $\pm$    0.0018 &     0.0610 $\pm$
		0.0012 &      0.64 $\pm$      0.22 \\

		&	$115$--$140$&   3.30 $\pm$      0.16 &      0.56 $\pm$     0.06 &     0.0301 $\pm$    0.0018 &     0.0735 $\pm$
		0.0010 &       1.09 $\pm$      0.21 \\
		
		&	$140$--$200$&  3.93 $\pm$      0.14 &      0.60 $\pm$     0.06 &     0.0356 $\pm$    0.0015 &     0.0831 $\pm$
		0.0010 &       1.18 $\pm$      0.12 \\

		\hline 
		
	\end{tabular}
	\label{tab:tab10}
\end{table*}


\bsp	
\label{lastpage}
\end{document}